\documentclass{svjour3}                     
\usepackage{natbib}
\usepackage{graphicx} 
\usepackage{times} 

\newcommand{\aap}{    {\rm Astron. Astrophys.}\ }

\newcommand{\apj}{    {\rm Astrophys. J.}\ }
\newcommand{\apjl}{   {\rm Astrophys. J. Lett.}\ }

\newcommand{\solphys}{{\rm Solar Phys.}\ }

\def\newblock{\rm }

\begin{document}
\title{Photospheric and Subphotospheric
Dynamics of Emerging Magnetic Flux}


\author{A.~G. Kosovichev}
 \institute{Stanford University,
Stanford, CA 94305, USA\\
\email{AKosovichev@solar.stanford.edu}}
\date{Received: date / Accepted: date}

\maketitle




\begin{abstract}
Magnetic fields emerging from the Sun's interior carry information
about physical processes of magnetic field generation and
transport in the convection zone. Soon after appearance on the solar surface
the magnetic flux gets concentrated in sunspot regions and causes
numerous active phenomena on the Sun. This paper discusses some
properties of the emerging magnetic flux observed on the solar
surface and in the interior. A statistical analysis of variations of the tilt
angle of bipolar magnetic regions during the emergence shows that
the systematic tilt with respect to the equator (the Joy's law) is most likely
established below the surface. However, no evidence of the dependence of
the tilt angle on the amount of emerging magnetic flux, predicted by the rising magnetic
flux rope theories, is found. Analysis of surface plasma flows in
a large emerging active region
reveals strong localized upflows and downflows at the initial phase
of emergence  but finds no evidence for large-scale flows
indicating future appearance a large-scale magnetic structure. Local helioseismology
provides important tools for mapping perturbations of the wave speed and mass
flows below the surface. Initial results from SOHO/MDI and GONG reveal
strong diverging flows during the flux emergence, and also localized converging
flows around stable sunspots. The wave speed images obtained during the process
of formation of a large active region, NOAA 10488, indicate that
the magnetic flux gets concentrated in strong field structures just below
the surface. Further studies of magnetic flux emergence
require systematic helioseismic observations from the ground and space, and realistic MHD
simulations of the subsurface dynamics.

\keywords{Solar magnetism \and Magnetic flux \and Active regions \and Sunspots \and Helioseismology}

\end{abstract}

\section{Introduction}
 The current paradigm is that the
solar magnetic fields are generated by a dynamo action deep in the
convection zone, presumably, at the bottom, in a thin rotational
shear layer called tachocline. In the tachocline the solar
differential rotation changes from the differential rotation of the
convection zone to a solid-body rotation of the radiative core. Most
of the tachocline is located in a convectively stable zone mixed by
convective overshoot \citep{Kosovichev:1996}. The combination of
strong shearing flows and stability makes possible generation
and storage of magnetic field in the tachocline \citep{Parker:1993}.

When the magnetic field is sufficiently strong it becomes buoyant and
emerges in the form of toroidal flux ropes ("$\Omega$-loops")
oriented in the East-West direction forming bipolar active regions
on the surface. The rising magnetic loops are affected by the
Coriolis force, which induces retrograde flows, directed from the
leading part of the toroidal tube towards its following part. The
Coriolis force acting on these flows causes deflection of the flux
tubes to higher latitudes and also a tilt with respect to the equator.
However, observations show that the magnetic flux emerges mostly
at mid and low latitudes, and calculations demonstrate that to
explain this the magnetic field generated at the bottom of the
convection zone must be very strong, 60--160 kG  \citep{DSilva:1992,DSilva:1994,Parker:1994}.
This is significantly higher than the field strength estimated from energy
equipartition with convective motions. Whether this is possible is
under debate \citep[e.g.][]{Schuessler:2005}.
\begin{figure}
\begin{center}
\includegraphics[width=\linewidth]{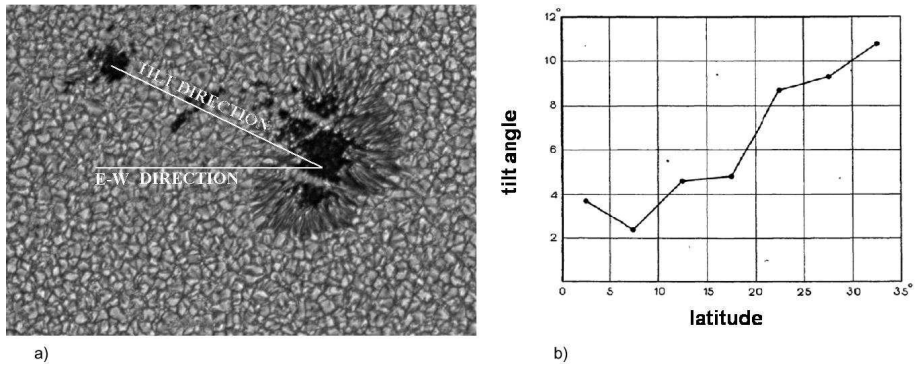}
 \end{center}
   \caption{a) Illustration of the Joy's law (courtesy of D.Hathaway). b)
   The tilt angle as a function of latitude \citep{Hale:1919}.}
    \label{Fig1}
\end{figure}

In addition, the Coriolis force causes a tilt in the orientation of
emerging flux tubes with respect to the equator. This phenomenon
is well-known as the Joy's law. Statistical studies by A.~H.~Joy
\citep{Hale:1919} of long series of sunspot drawings showed that the
following spot of a bipolar group tend to appear farther from the
equator than the preceding spot, and that the higher the latitude the
greater the inclination of the axis to the equator
Fig.~\ref{Fig1}. The tilt of bipolar magnetic groups appears
only statistically. The orientation of individual active regions may
vary quite significantly. This effect is important  for
understanding the process of magnetic flux emergence. Also, it is
a key element of solar-cycle theories \citep[e.g.][]{Wang:1991}.

Theories of rising magnetic flux tubes seem to explain the tilt
\citep{Schmidt:1968}. In these theories, the tilt angle is
determined by the torque balance between magnetic tension and
the Coriolis forces, and, thus, depends on both the amount of magnetic
flux and the emergence latitude. \citet{Fisher:1995} and
\citet{Caligari:1995} found from an analysis of sunspot group data that
the theoretically predicted flux dependence of the tilt angle was consistent with
the data. \citet{Tian:2003}
updated these results using magnetic fluxes instead of polarity
separations for a more direct comparison with the theory.
Their results showed a less clear agreement between the rising flux tube theories and the
observations. These theories also predict that after the rising phase
the magnetic flux tubes become stationary, and since the Coriolis
force disappears the tilt should decrease under the action of
magnetic tension in the East-West direction. However, the tilt of bipolar groups does not
disappear after the emergence \citep{Howard:2000}. Moreover, there
is a tendency of the tilt angle to rotate towards the averaged value
defined by the Joy's law.
This cannot be explained by the Coriolis effect, and is probably
related to a complicated interaction between the magnetic structures
and flows below the surface. An alternative explanation suggested by
\citet{Babcock:1961} is that the tilt is due to the spiral orientation of the magnetic
field lines below the surface, wrapped around the Sun by the differential rotation. However, his mechanism
assumes that the toroidal field is generated close to the surface,
not in the tachocline.

In general, understanding of the observed properties of emerging
magnetic flux is closely related to the depth of the main dynamo process in
the Sun. While the modern theories assume that the solar dynamo
operates in the tachocline there is no convincing observational
evidence to support this. Also, there are theoretical difficulties.
The pros and cons of the dynamo mechanisms operating in the
tachocline and in the bulk of the convection zone, or perhaps even
in the near-surface shear layer, are discussed by
\citet{Brandenburg:2005}. From the observational point of view,
helioseismology observations for the whole solar cycle from the Michelson Doppler Imager (MDI) instrument on Solar and Heliospheric
Observatory (SOHO) \citep{Scherrer:1995} and Global Oscillation Network Group (GONG)
\citep{Harvey:1996}
do not provide a convincing evidence for
solar-cycle variations of the solar rotation rate in the tachocline
\citep{Howe:2007}.  Such variations are expected because of the back reaction of the
strong dynamo-generated magnetic fields on the turbulent Reynolds
stresses and, hence, on the differential rotation (which is maintained by
the Reynolds stresses). Moreover, the
comparison of the rotation rate of long-living complexes of activity
(which are the sources of repeated flux emergence) with the internal
differential rotation deduced from helioseismology showed that the
roots of the complexes of activity are probably located in the
near-surface shear layer \citep{Benevolenskaya:1999}. Determination of the
depth of the solar dynamo is probably one of the most important problems
of solar magnetism.

Observations show that emerging magnetic flux plays an important role
in initiation of solar flares and coronal mass ejections. Thus, it
is important to develop predicting capabilities for flux emergence.
This problem can be addressed by helioseismology but initial attempts
to detect the magnetic flux in the interior before it becomes
visible on the surface showed that this is difficult because of the
high emergence speed in the upper 20 Mm \citep{Kosovichev:2000}.
Thus, it is important to investigate large-scale flow patterns,
which may provide indication of the flux emergence and development
of large magnetic regions in the interior.

In general, investigation of emerging magnetic flux includes the
following questions:
\begin{itemize}
\item
How deep is the source of emerging magnetic flux?
\item
Does emerging magnetic flux become disconnected from the source?
\item
Why does magnetic flux tend to emerge in the same areas, forming
complexes of activity?
\item
What is the plasma dynamics associated with emerging flux?
\item
How does emerging flux interact with the existing magnetic fields,
and triggers flares and CMEs?
\item
Can we predict emerging magnetic flux before it become visible on
the surface?
\item
Can we predict evolution of active regions?
\end{itemize}

This paper discusses some of these questions and presents recent
results of investigation of surface and subsurface characteristics
of the magnetic flux emergence process obtained from SOHO/MDI
 and GONG. In particular, I discuss
a new analysis of the Joy's
law for the emerging flux, dynamics of
the photospheric plasma prior and during the flux emergence, methods
and results of acoustic tomography of wave-speed perturbations and
mass flows below the visible surface, and also future observational
projects and perspectives.

\section{Observations of Emerging Magnetic
Flux in the Photosphere}
\subsection{Joy's Law and Magnetic Flux Transport}
\begin{figure}
\begin{center}
\includegraphics[width=\linewidth]{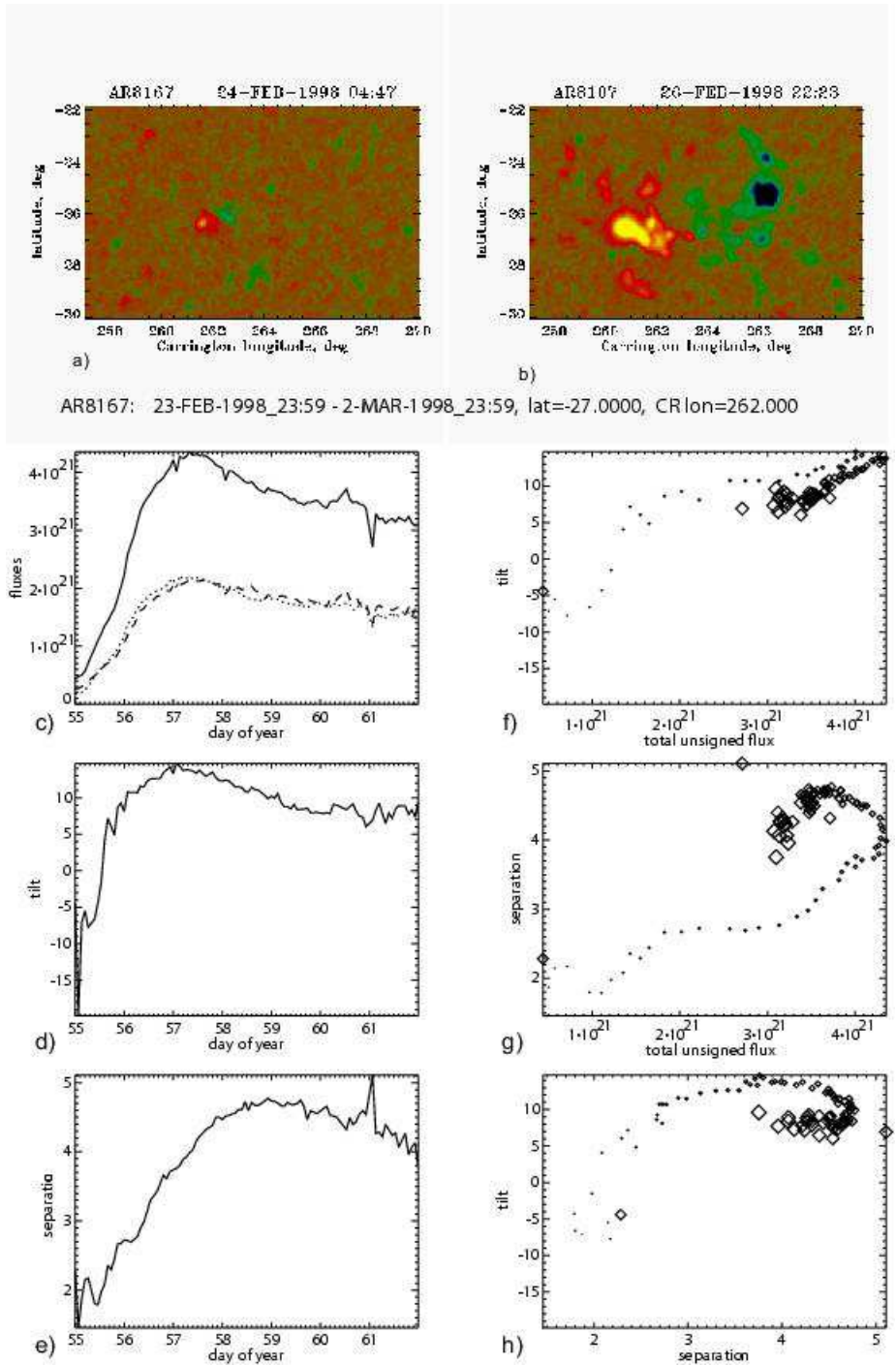}
 \end{center}
   \caption{Magnetograms of active region NOAA 8167: a) at the beginning
of flux emergence and b) at the end of emergence. The green and blue colors
show the negative polarity; the red and yellow colors show the positive
polarity. The
   evolutions of: c) magnetic fluxes in Mx (dashed curve - positive polarity,
dotted curve - negative polarity, the solid curve - the total unsigned flux);
d) the tilt angle (in degrees);  e) the separation between the polarities
(in heliographic degrees). The relationships between: f) the tilt angle and
total unsigned flux,
   g) the polarity separation and the total flux, h) the tilt angle and the separation.
The symbol size is proportional to time from the start of emergence.}
    \label{Fig2}
\end{figure}
The tilt of bipolar sunspot groups with respect to the equator
(the Joy's law) is one of the fundamental properties of solar magnetism.
This phenomenon is closely related to the dynamo mechanism and the
process of flux emergence. The key question is whether the tilt is caused
by the Coriolis force acting on magnetic flux tubes radially moving
from the bottom of the convection zone \citep{Schmidt:1968}, or
it reflects the orientation of subsurface magnetic field lines
stretched by the differential rotation \citep{Babcock:1961}, or
it is created by subphotospheric shearing flows after the emergence
\citep{Howard:1996a}.  Previous studies of the Joy's law were based
on daily white light images or magnetograms of sunspot group. These data did not
have sufficient temporal resolution to investigate variations of the tilt angle during the flux
emergence process.

Using a series of 96-min cadence magnetograms from SOHO/MDI,
\citet{Kosovichev_Stenflo:2008} attempted to investigate the tilt angle
and its statistical relationships to the region latitude, the amount of
emerging flux, the emergence rate and the separation between the magnetic
polarities. The magnetograms obtained almost uninterruptedly for
almost the whole solar cycle, from May 1996 until October 2006, have been
analyzed. During this period the MDI instrument on SOHO observed
more than 2000 active regions, and 715 active regions, which
emerged within 30 degrees from the central meridian, are selected
for this study.

The analysis method is pretty straightforward. Each active region
is remapped into the heliographic coordinates.
The tilt angle and the separation between the magnetic polarities
are calculated for their centers of
gravity. The period of the growth of the
total magnetic flux  (lasting usually 2-3 days) is divided into 5
intervals, and the statistical relations are calculated for each
interval separately, and for the whole emergence phase.

A typical example of this data analysis  is shown in Fig.~\ref{Fig2}
for active region NOAA 8167 emerged in the Southern
hemisphere at about $26^\circ$ latitude. The magnetic flux of both
polarities rapidly and simultaneously increased  and reached a maximum
during the first 2 days after the initial appearance of the bipolar
region on the surface (Fig.~\ref{Fig2}c). The tilt angle
(Fig.~\ref{Fig2}d) shows rapid variations at the beginning of
emergence, reaches a maximum of about $15^\circ$ and then stabilizes at
about $7-8^\circ$. The separation between the polarities
(Fig.~\ref{Fig2}e) increases during the emergence phase and
continues after the total flux reaches the maximum. It starts
decreasing as the active region decays. Figures~\ref{Fig2}f-h show
the relationships among these properties. In this example the tilt
angle is established rather quickly during the emergence in
accordance with the Joy's law (the blue polarity is closer to the
equator than the red polarity in Fig.~\ref{Fig2}b). However, the
title angle varies significantly in our sample of emerging active
regions, and the Joy's law holds only statistically.

\begin{figure}
\begin{center}
\includegraphics[width=0.9\linewidth]{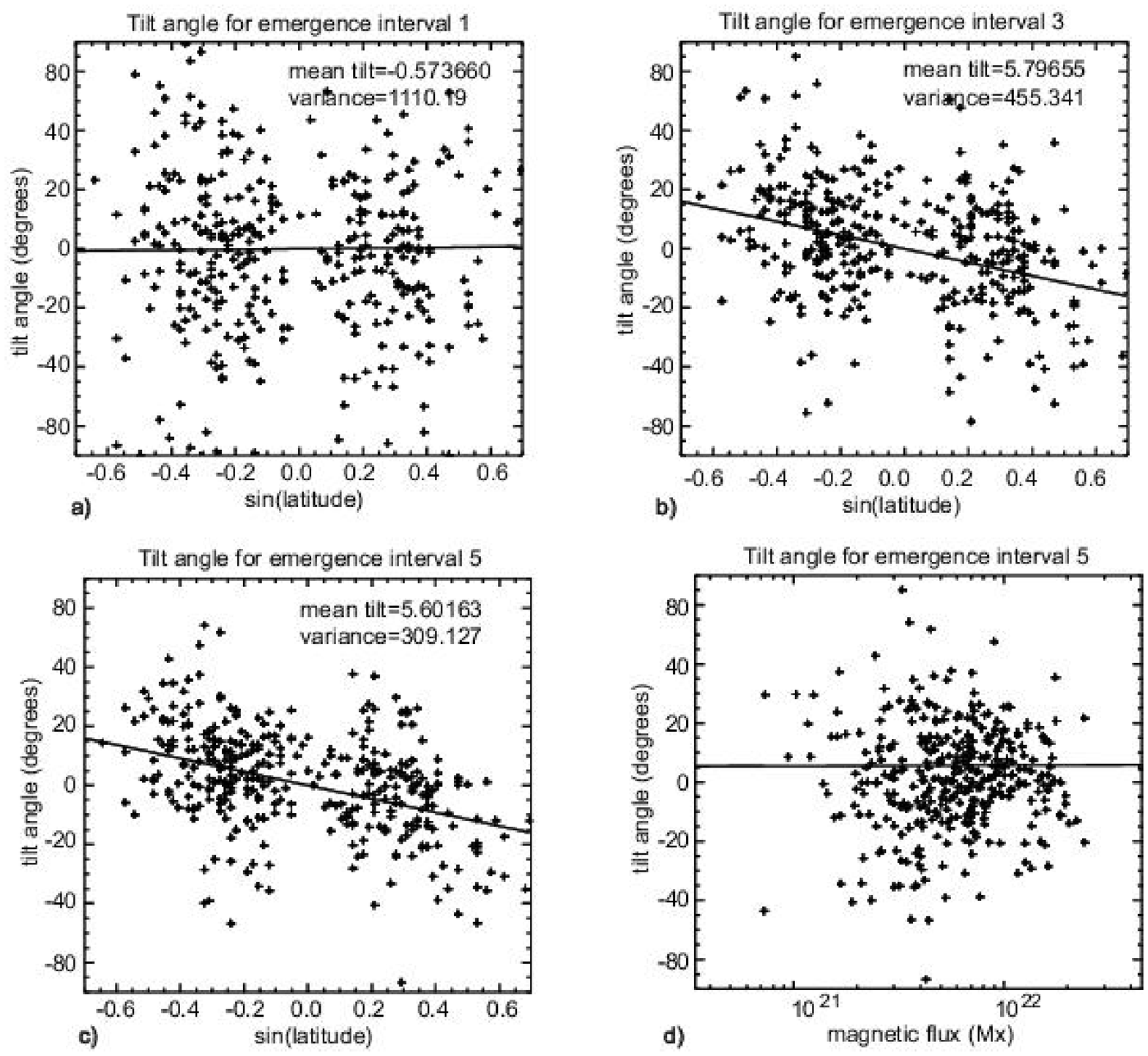}
 \end{center}
   \caption{The distribution of the tilt angle with sine latitude
at the beginning of emergence (a), at the middle of the emergence
interval (b), and at the end of emergence (c); the distribution of
the tilt angle with the total magnetic flux at the end of
emergence.}
    \label{Fig3}
\end{figure}

Figures~\ref{Fig3}a-c show the distributions of the tilt angles with
latitude for three periods of emergence (the total flux growth): the
initial appearance (emergence interval 1), the mid interval (3) and
at the end of emergence (interval 5). It appears that at the
beginning of emergence the tilt angle is randomly distributed, and
the mean tilt angle is about zero (Fig.~\ref{Fig3}a). However, at
the middle of the emergence period the distribution of the tilt
angle clearly follows the Joy's law (Fig.~\ref{Fig3}b) with the
latitudinal dependence and the mean tilt angle of about 6 degrees.
At the end of the emergence period the Joy's law distribution
becomes more pronounced as the variance of the deviation from the
linear dependence on the sine latitude decreases (Fig.~\ref{Fig3}c).
However, these data show no significant correlation between the tilt
angle and the total magnetic flux at the end of emergence
(Fig.~\ref{Fig3}d).

These results show that the tilt of bipolar magnetic regions becomes
statistically significant during the emergence process. This means
that the tilt is established in subsurface layers. Among the
mechanisms suggested to explain the observed tilt are: the spiral
orientation of the subsurface toroidal field lines wrapped around
the differential rotation \citep{Babcock:1961}, the effect of the
Coriolis  force acting on the flux tubes moving from the bottom of
the convection zone \citep{Schmidt:1968}, and large-scale subsurface
motions associated with the differential rotation and meridional
circulation \citep{Howard:1996a}. The most popular explanation that
the tilt is caused by the Coriolis force acting on the flows inside
an emerging flux rope was questioned by \citet{Howard:1996a} who
investigated variations of the tilt after the emergence and found
that the tilt angle moves towards the Joy's law orientation instead
of relaxing to the East-West direction as expected from this theory
when the radial flux rope motion stops (and, thus, the Coriolis force
vanishes).  In addition, our results do not show a significant
dependence of the tilt on the magnetic flux, predicted by the
Coriolis force theories \citep[e.g.][]{Fan:1994}.
\citet{Howard:1996a} suggested that the tilt angle may be
established after the emergence due to the action of the depth
dependent differential rotation and meridional flow. However, our
results indicate that the bipolar magnetic flux regions emerge already
tilted in accordance with the Joy's law. The helioseismology results show that the emerging
magnetic structures propagate very fast in the upper convection zone (Sec.~3.2), and thus
do not support the Howard's idea. Perhaps, we should go back
to the Babcock's mechanism that the tilt is caused by the spiral
structure of the subsurface toroidal flux tubes.

\begin{figure}
\begin{center}
\includegraphics[width=0.7\linewidth]{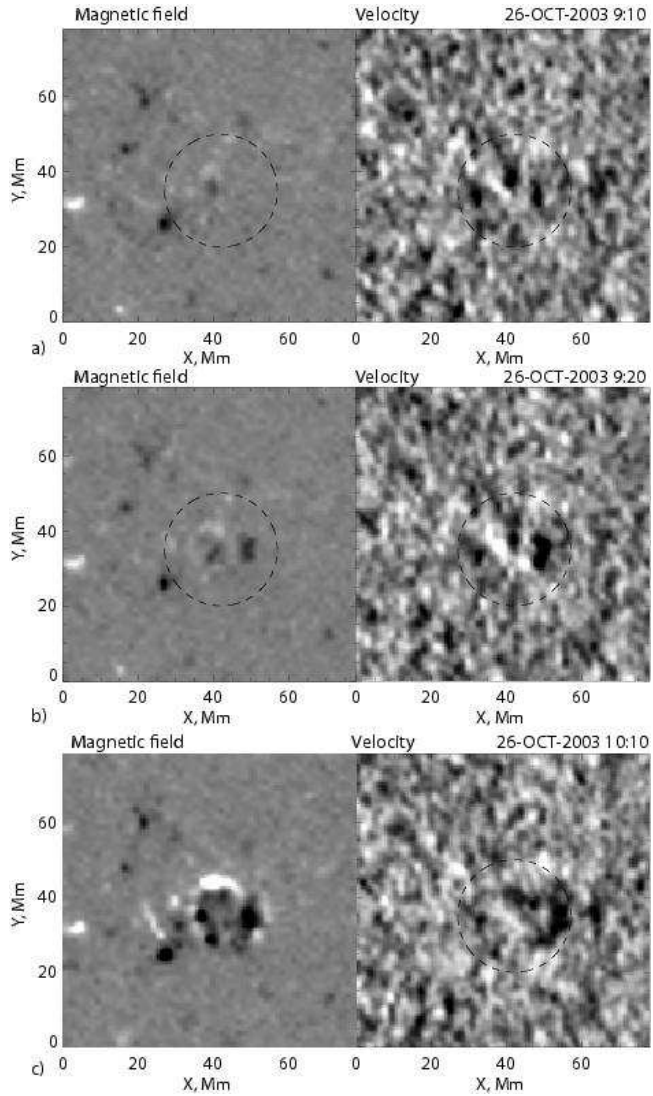}
   \caption{Maps of the line-of-sight magnetic field (left panels)
and Doppler velocity (right panels) on the solar surface obtained
from SOHO/MDI at the beginning of emergence of AR 10488, October 26,
2008, a) 7:30 UT, b) 7:40 UT, c) 7:50 UT. The range of the magnetic
field strength is [-180 G, 180 G]. The range of the Doppler velocity
is [-600 m/s, 600 m/s]. The dark color shows upflows, and the white
color shows downflows. The dashed circle outlines the area of the
initial magnetic flux emergence.}
    \label{Fig4}
 \end{center}
\end{figure}

\subsection{Mass Flows}

One can expect that when a large magnetic flux rope emerges on the solar
surface it drives significant upflows and outflows, which may be
detectable when the flux rope is still below the surface.
Figure~\ref{Fig4} shows the magnetograms and Dopplergrams obtained
from SOHO/MDI on October 26, 2003, during the initial emergence of
AR 10488, which later grew in to one of the largest active regions of Solar
Cycle 23. These data reveal strong localized upflows (dark features
in the velocity images) in the places of the initial magnetic flux
growth at 9:10~UT (Fig.~\ref{Fig4}a). Ten min later the upflow velocity
reaches a peak of about 800 m/s, and the magnetic flux starts
appearing on the surface (Fig.~\ref{Fig4}b). The strong upflow is
mostly concentrated in the leading part of the emerging flux. In the
following part, the data show the Doppler shift of the opposite sign
corresponding to downflows. After the appearance of the magnetic
flux and its initial growth we observe the similar flow pattern with
upflows in the leading part, but the velocity amplitude decreases
(Fig.~\ref{Fig4}c).

\begin{figure}
\begin{center}
\includegraphics[width=0.7\linewidth]{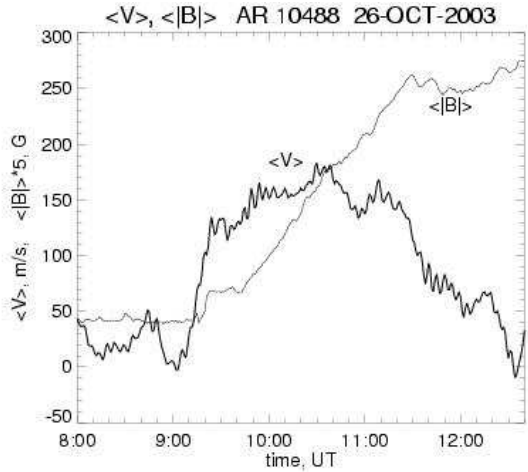}
 \end{center}
   \caption{The mean Doppler velocity and the total magnetic flux as a function
of time in the region of the initial emergence of AR 10488. The
positive velocity values correspond to the plasma motions towards
the observer (upflow).}
    \label{Fig5}
\end{figure}

The evolution of the mean Doppler velocity and the total magnetic
flux in  the area of emergence (Fig.~\ref{Fig5}) shows a sharp rise
of upflows, which continue to be strong for about 2 hours during the
initial emergence. The mean velocity reaches $\sim 150$ m/s, well above
the mean velocity fluctuations of about 50 m/s in similar-sized quiet-Sun
regions. After the initial emergence phase, we do not observe significant upflows
despite the continuing growth of the active regions.

The strong photospheric plasma flows associated with the magnetic
flux emergence of this active region have been also detected by
\citet{Grigoriev:2007}. It is unclear if such surface flows
are typical for emerging magnetic flux and if their strength
corresponds to the amount of emerging flux. \citet{Pevtsov:2006}
studied plasma flows in fifteen emerging active regions using
Dopplergrams, magnetograms, and white light observations from
SOHO/MDI. They observed no consistent plasma flows at the future
location of an active region before its emergence.

Also, in the case of AR10488 there is a systematic upflow in the
leading magnetic polarity and a downflow in the following polarity.
This can interpreted as flows inside the emerging flux tubes, driven
by the Coriolis force \citep[e.g.][]{Fan:1994}. However,
\citet{Pevtsov:2006} found the asymmetric flows only in three active
regions. In two regions, flows are directed from the following to
leading polarity, and in one region material flows from the leading
to the following polarity. Thus, more detailed statistical studies
of the flow dynamics in emerging magnetic flux are necessary.

\section{Investigations of Emerging Flux by Helioseismology}

Methods of local helioseismology developed in recent years allow us
to probe physical conditions of the solar plasma below the surface
and detect the wave-speed structures and mass flows associated with
the emerging magnetic flux.

\subsection{Method of Time-Distance Helioseismology}
\begin{figure}
\begin{center}
\includegraphics[width=\linewidth]{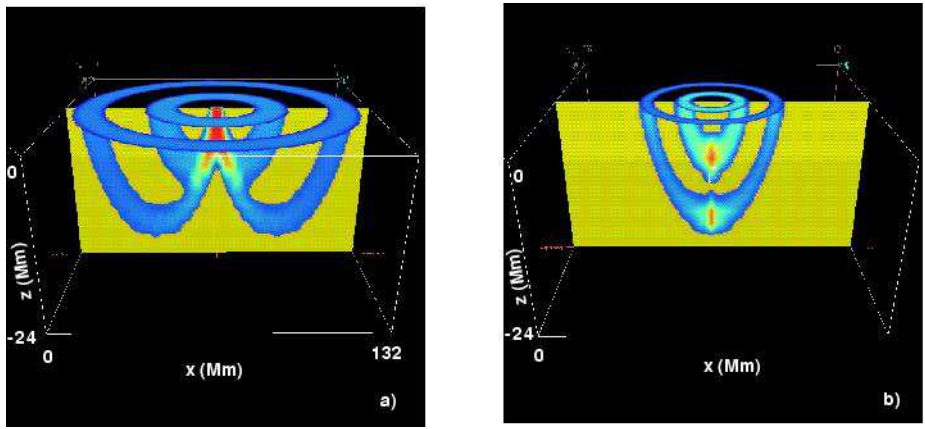}
 \end{center}
   \caption{Illustration of time-distance sensitivity kernels calculated
in the ray-path approximation: a) surface-focusing scheme; b)
deep-focusing scheme.}
    \label{Fig6}
\end{figure}

\begin{figure}
\begin{center}
\includegraphics[width=0.7\linewidth]{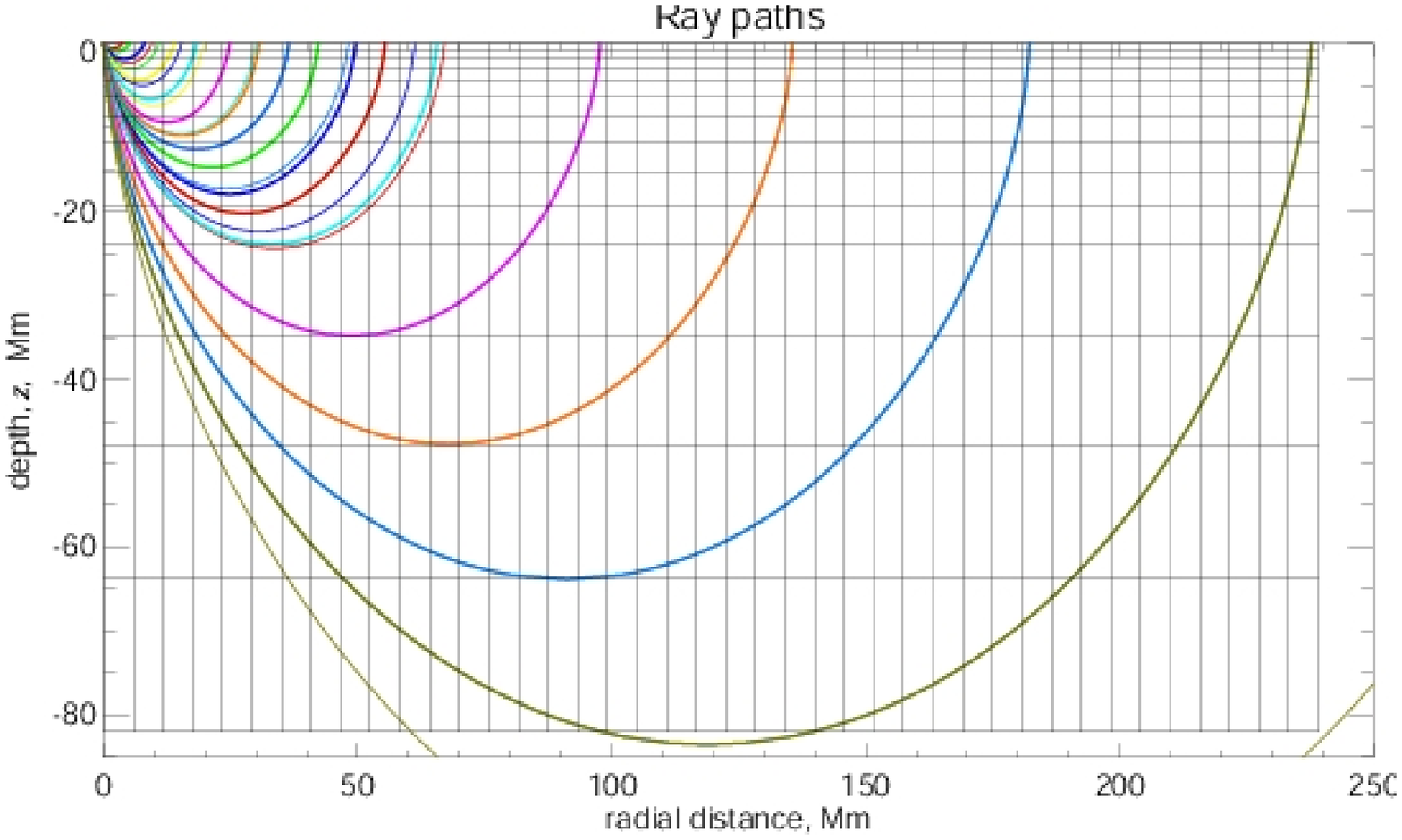}
 \end{center}
   \caption{A vertical cut through the 3D data inversion grid and a sample of acoustic
ray paths.}
    \label{Fig7}
\end{figure}
Time-distance helioseismology measures travel times of acoustic
waves propagating to different distances, and uses these
measurements to infer variations of the wave speed along the wave
paths. Turbulent convection excites acoustic waves which propagate
deep into the solar interior. Because the sound speed increases with
depth these waves are refracted  and come back to the solar surface.
The wave speed depends on the temperature, magnetic field strength and
flow velocity field in the region of the wave propagation. By
measuring reciprocal travel times of acoustic waves propagating
along the same ray paths in opposite directions, and then taking the
mean and the difference of these travel times, it is possible to
separate the flow velocity (advection) effect  from temperature and
magnetic field perturbations \citep{Kosovichev:1997}. However, in
order to disentangle the contributions of temperature variations and magnetic
field to the mean travel times it is necessary to measure the
travel-time anisotropy, and this has not been accomplished.
Therefore, the current helioseismic results represent maps of
sub-photospheric variations of the magneto-acoustic wave speed
and flow.

The travel times are typically measured from a cross-covariance
function of solar oscillation signals for various distances and time
lags. When for a given distance the time lag corresponds to the
propagation time of acoustic waves for this distance, a
wavepacket-like signal appears in the cross-covariance function. The
cross-covariance plotted as a function of the distance and the time lag
displays a set of ridges formed by the wave-packet signals,
representing an analog of a solar ``seismogram''. Since the solar
oscillations are stochastic it is necessary to use the oscillation
signals at least 2--8 hours long and also average them over some
surface (typically, circular) areas in order to obtain a sufficient
signal-to-noise ratio. Then, the travel times are determined by
fitting a wavelet to this function  \cite[e.g.][]{Kosovichev:1997},
or by measuring displacement of the ridges \citep{Gizon:2002}. Two
general observing schemes, so-called `surface-focusing' and
`deep-focusing',  have been used in the travel-time measurements. In
the surface-focusing scheme the travel times are measured for
acoustic waves traveling between a central point and surrounding
annuli. In the deep-focusing scheme, the travel times are measured
for the acoustic wave packets traveling between the opposite parts
of the annuli, the ray paths of which cross each other in a point
located below the surface.

The relationship between the observed travel-time  variations and
the internal properties of the Sun is given by so-called sensitivity
kernels (illustrated in Fig.~\ref{Fig6} for both surface- and deep focusing) through integral equations.
These integral equations are solved by standard mathematical
inversion techniques such as Least Square QR Decomposition (LSQR)
and Multi-Channel Deconvolution
(MCD) \citep*{Kosovichev:1996,Jensen:2001,Couvidat:2006}. The sensitivity
functions are calculated using a ray theory or more complicated wave
perturbation theories, e.g., the Born approximation, which takes
into account the finite wave-length effects. These theories can also
take into account stochastic properties of acoustic sources
distributed over the solar surface \citep*{Gizon:2002,Birch:2004}.

The vertical structure of the computational grid and a sample of
acoustic ray paths, used in this paper, are illustrated in
Figure~\ref{Fig7}. The travel times are measured for waves
traveling between a central location and surrounding annuli with
different radial distances from the central point. The width of the
annuli is larger for larger distances in order to improve the
signal-to-noise ratio. A set of 17 annuli covering the distance
range from 0.54 to 24.06 heliographic degrees (from 6.5 to 292 Mm)
was used. The acoustic waves traveling to these distances sample the
Sun's interior up to the depth of 95 Mm. The central locations of
the time-distance measurements are chosen on a uniform $256\times
256$ grid with the grid step of 2.9 Mm. A total of $1.1\times 10^6$
travel time measurements are made to obtain each wave-speed image
of the interior. For the flow velocity, the number of the
measurements is three times larger, because in this case in addition
to the travel times for a whole annulus it is necessary to measure
also the travel times for waves traveling North--South and
East--West. This is done by dividing each annulus into four sectors.
A part of the inversion grid and the ray paths of the acoustic waves
are illustrated in Fig.~\ref{Fig7}). The horizontal step in this figure
is twice as large as the step of the travel-time measurements.
The vertical grid, 82 Mm deep, is
non-uniform with the step size increasing with depth, from 0.7 Mm at
the top to 17 Mm at the bottom. The inversion procedure for all
three components of flow velocity is described by
\citet{Kosovichev:1997}.

\subsection{Tomographic Imaging of Wave-Speed Perturbations}

\begin{figure}
\begin{center}
\includegraphics[width=0.8\linewidth]{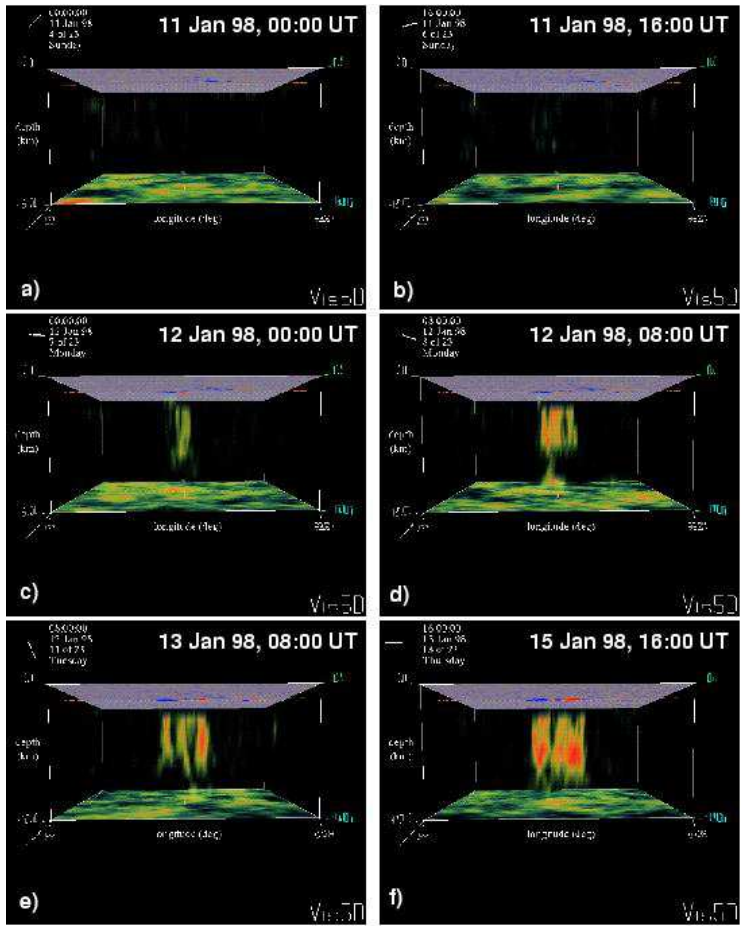}
 \end{center}
   \caption{The wave-speed perturbation in the emerging active region,
   NOAA 8131. The horizontal size of the box is approximately 38 degrees (460 Mm),
the vertical size is 18 Mm.
The panels on the top are MDI magnetograms showing the surface magnetic field of
positive (red) and negative (blue) polarities.
The perturbations of the wave speed are approximately in the range from -1 to +1 km/s.
The positive
variations are shown in red, and the negative ones in blue.}
    \label{Fig8}
\end{figure}

\begin{figure}
\begin{center}
\includegraphics[width=0.6\linewidth]{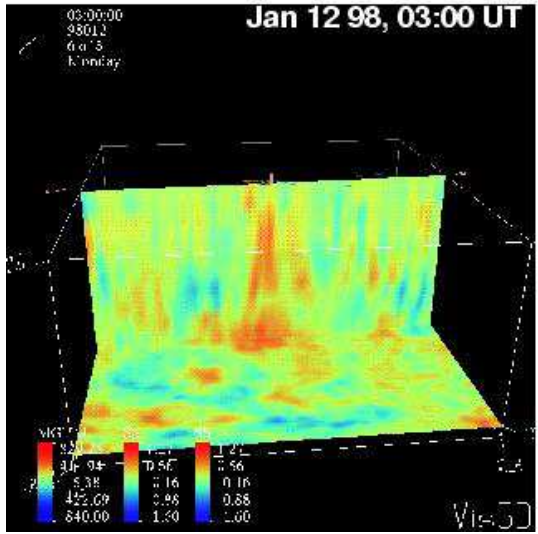}
 \end{center}
   \caption{Image of the magneto-acoustic wave speed in an emerging active region (AR 8131) in the
solar convection zone obtained from the SOHO Michelson Doppler Imager
(MDI) data on January 12, 1998, from 02:00 to 04:00 UT, using
time-distance helioseismology.  The horizontal size of the box is
approximately 560 Mm, and the depth is 18 Mm. The (mostly) transparent
panel on the top is an MDI magnetogram showing the surface magnetic
field of positive (red) and negative (blue) polarities stronger than
200 Gauss. The vertical and bottom panels show perturbations of the
wave speed which are approximately in the range from -1.3 to
+1.3 km/s. The positive variations are shown in red, and the negative
ones in blue.  A large active region formed at this location within
a day after these observations.
}
    \label{Fig9}
\end{figure}

Figure~\ref{Fig8} shows the results for the emerging active region, NOAA 8131,
of January, 1998, obtained by \citet{Kosovichev:2000}.
This was a high-latitude region of the new solar cycle which began in 1997.
The distribution of the wave speed variations
in a vertical cross-section in the region of the emerging flux
and in a horizontal plane at a depth of 18 Mm are shown for six 8-hour
consecutive intervals. The perturbations
of the magnetosonic speed shown in this figure are associated with
the magnetic field and temperature variations in the emerging magnetic ropes
and in the surrounding plasma.
The panel (a) shows no significant variations in the
region of  emergence, which is at the middle of the vertical
plane. The MDI magnetogram shown at the top indicates
only very weak magnetic field above this region.
The panel (b) shows a slight positive perturbation associated with
the emerging region.
During the next 8 hours
(panel c) the perturbation becomes stronger and occupies the whole range
of depths and continue to increase.
These results show that the emerging flux propagates very quickly
through the upper 18 Mm of the convection zone.
We have also analyzed the data for 2-hour intervals at the start
of emergence from 2:00~UT to 4:00~UT, January 12, 1998, (Fig.~\ref{Fig9}) and
concluded that the emerging flux propagated through
the characteristic depth of 10 Mm in approximately 2 hours. This gives an
estimate of the speed of emergence $\approx 1.3$ km/s. This speed
is similar to the speed predicted by the theories of emerging
flux ropes. The typical amplitude
of the wave-speed variation in the emerging active region is about 0.5 km/s.
After the emergence we observed a gradual increase of the
perturbation in the subsurface layers, and the formation of sunspots (Fig.~\ref{Fig7}d-f).
The observed development of the active region seems to suggest that the sunspots
are formed as a result of the concentration of magnetic flux close to the
surface.

\begin{figure}
\begin{center}
\includegraphics[width=\linewidth]{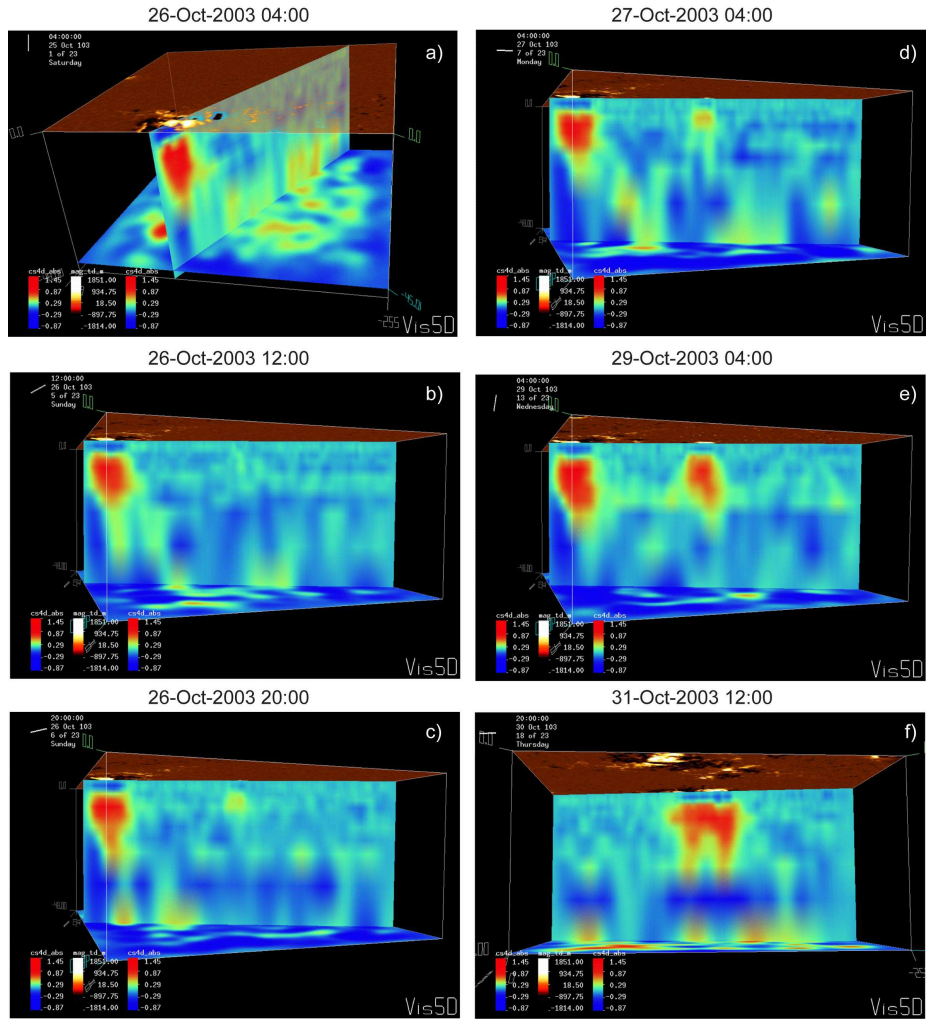}
 \end{center}
   \caption{Subsurface magnetosonic wave-speed structures of the large complex of
activity of October--November 2003, consisting of active regions
NOAA 10486 (in the left-hand part of the images), and 10488
(emerging active region in the middle). Red color shows positive
wave-speed variations relative to the quiet Sun; the blue color
shows the negative variations, which are concentrated near the
surface. The upper semi-transparent panels show the corresponding
MDI magnetograms; the lower panel is a horizontal cut 48 Mm deep.
The horizontal size is about 540 Mm. The vertical cut goes through
both active regions, approximately in the North--South direction
crossing the equator, except the image in the right bottom panel,
(f), where it goes only through AR 10488 in the East--West
direction.}
    \label{Fig10}
\end{figure}

\begin{figure}
\begin{center}
\includegraphics[width=\linewidth]{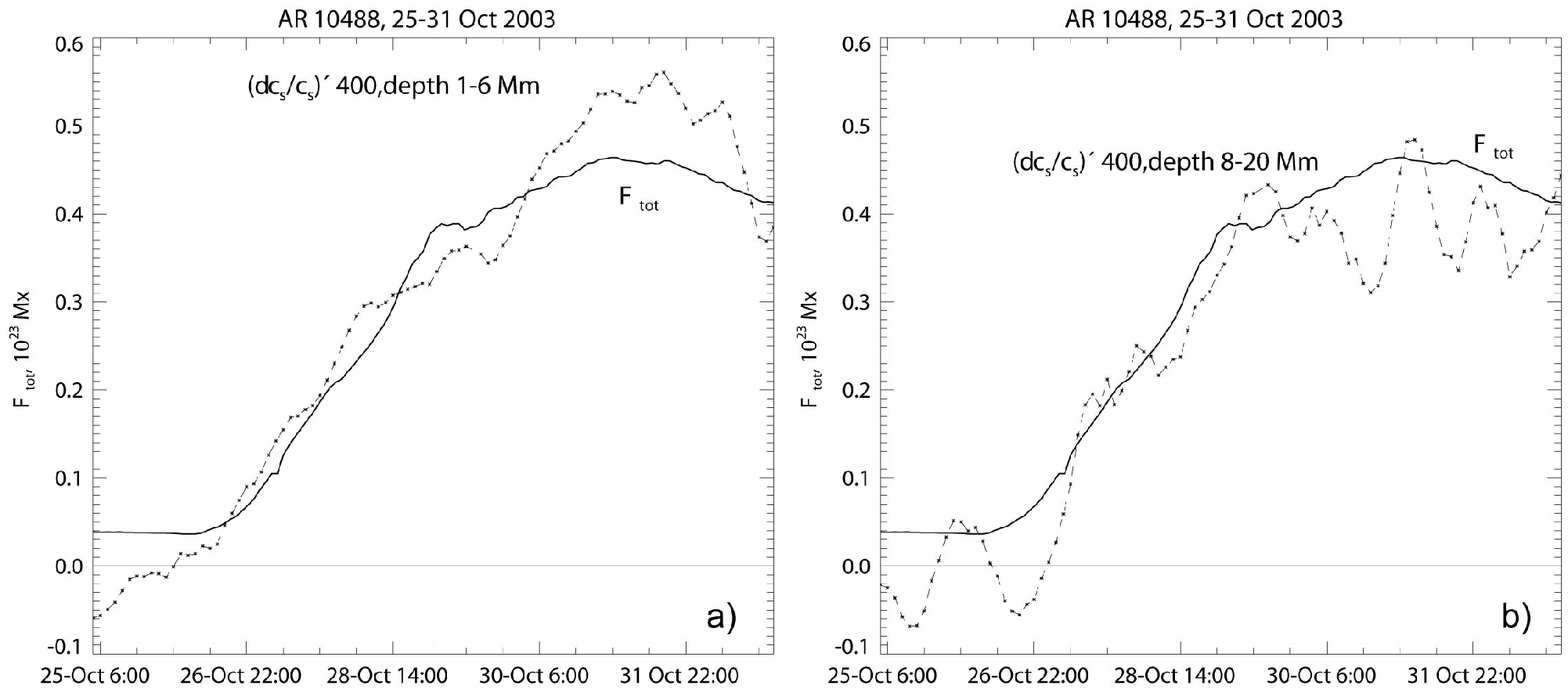}
 \end{center}
   \caption{Evolution of the total unsigned photospheric magnetic
flux (solid curve) and the mean relative wave-speed variation
(dotted curve with stars) at the depth of 1--6 Mm ({\it left}) and
8--20 Mm ({\it right}) in the region of the flux emergence of AR
10488.}
    \label{Fig11}
\end{figure}
Several large active regions emerged on the Sun in October--November
2003. This period represents one of the most significant impulses of
solar activity in solar cycle 23. During this period the MDI
instrument on SOHO was in the full-disk mode (``Dynamics program''),
taking full-disk Dopplergrams and line-of-sight magnetograms every
minute. Thus, it was able to capture the emergence of active regions
NOAA 10488, and also obtain data for two other large active regions,
10484 and 10486. The active region, 10488, emerged in the Northern
hemisphere (at $291\deg$ Carrington longitude and $8\deg$ latitude)
approximately at the same longitude as AR 10486, which emerged
earlier and had a very complex magnetic configuration resulting in
several strong flares. It is possible that
these two active regions had a common nest in the interior
\citep{Zhou:2007}.

Using the method of time-distance helioseismology, we have obtained
wave-speed and flow velocity maps for 8 days, 25--31 October, 2003, \citep{Kosovichev:2008}.
The maps are obtained using 8-hour time series with a 2-hour shift.
Total 96 wave-speed and flow maps were obtained. Fig.~\ref{Fig10}
shows a sample of the wave-speed images. The vertical cut through
these images (except the image of 31 Oct 2003, 12:00) is made
through both AR 10486 and 10488 in approximately the North-South
direction, and for the image of 31 Oct 2003, 12:00,
(Fig.~\ref{Fig10}f) it is made in the East-West direction.
The depth of the image box is 48 Mm, and
the horizontal size is about 540 Mm.

The results show that the first wave-speed signal below the surface
appeared in the image obtained on 26 October, 2003, for the time
interval centered at 12:00 UT (Fig.~\ref{Fig10}b). This is
slightly ahead of the growth of the total magnetic field flux, which
started to grow at about 20:00 UT (Fig.~\ref{Fig11}, left, solid
curve; however, the first magnetic field signal appeared
approximately at the same time). During the next 8 hours, between
12:00 and 20:00, the wave-speed perturbation rapidly grows, and is
most visible in the subsurface layers, about 10 Mm deep. In the
deeper interior, we do not detect a clear signal above the noise
level at this time. This may be because the relative perturbation in
these layers is too weak, and also may indicate that the formation
of magnetic flux concentrations starts in the subsurface layers.
During the next 8 hours the signal extends into the deeper layers
and continues to grow (Fig.~\ref{Fig10}d). The typical
two-layer structure with lower wave speed in the top 4--5 Mm, and
higher wave speed in the deeper layers is formed
\citep*{Kosovichev:2000,Jensen:2001,Couvidat:2006}. During the following 5 days of the MDI
observations, the wave-speed perturbation below the active region
becomes larger and stronger, and in the East-West direction it forms
a loop-like structure (Fig.~\ref{Fig10}f). This structure
can be traced to the depth of about 30 Mm, and then it is lost in
noise.

In Figure~\ref{Fig11} we compare the evolution of the total
(unsigned) magnetic flux of the active regions and the mean
wave-speed signal in the two depth intervals, 1--6 Mm and 8--20 Mm.
In both cases, the  signals correspond well to the evolution of the
surface magnetic flux. There is possibly an indication of a slight lead of
the wave-speed signal at 1--6 Mm, but there is no significant time lag.
At greater depths the noise level is higher, and it is even more
difficult to see the difference in the time evolution relative to
the surface magnetic flux.

These results show that the magnetic flux emerges very rapidly from
the interior, and that there is no significant (on the scale of few
hours) time difference between the evolution of the wave-speed
variations associated with the emerging active region and the
photospheric magnetic flux. There are indications that the process of the magnetic
field concentration, which forms the active region, first occurs
in the subsurface layers, and that then the active region grows because
of subsequent flux emergence in this area.
\subsection{Subsurface Flows}
\begin{figure}
\begin{center}
\includegraphics[width=\linewidth]{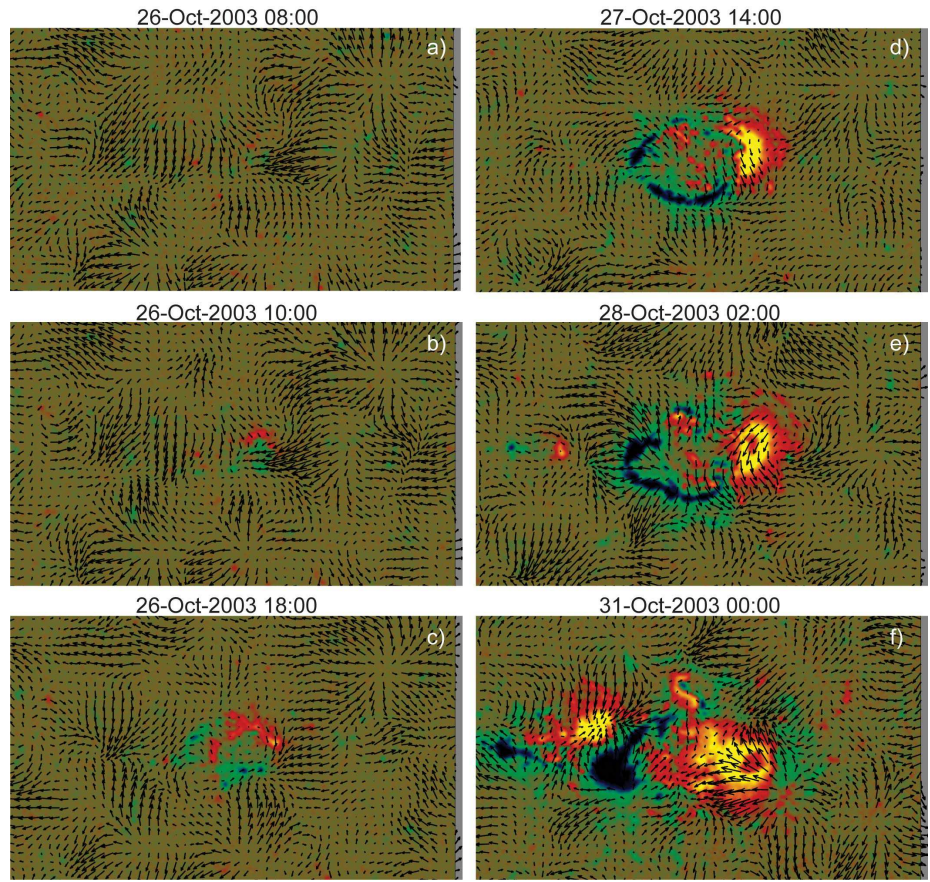}
 \end{center}
   \caption{Evolution of subsurface flows at the depth of 2 Mm below
the photosphere during the emergence and growth of AR 10488, on
26--31 October, 2003. The flow maps are obtained by the
time-distance technique using 8-hour time series of full-disk
Doppler images from SOHO/MDI. The maximum horizontal velocity is
approximately 1 km/s. The background image is the corresponding
photospheric magnetogram (red and blue areas show regions of
positive and negative polarity of the line-of-sight magnetic
field.)}
    \label{Fig12}
\end{figure}

\begin{figure}
\begin{center}
\includegraphics[width=\linewidth]{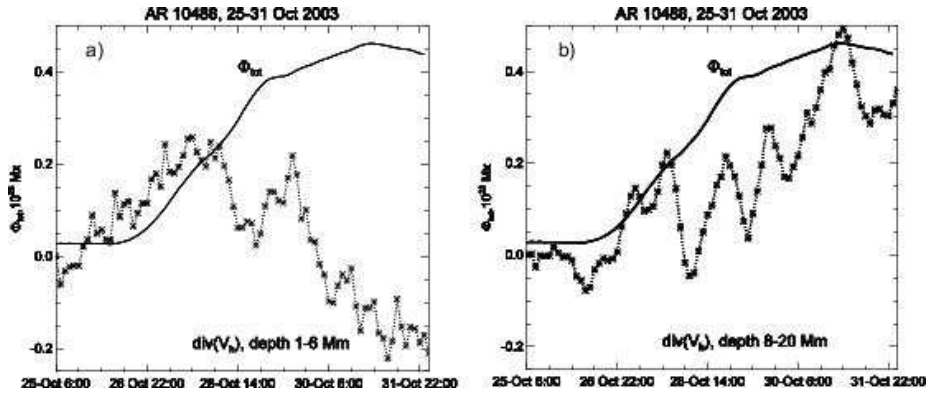}
 \end{center}
   \caption{The evolution of the total unsigned photospheric magnetic
flux (solid curve) and the mean divergence of the horizontal flow
velocity (dotted curve with stars) at the depth of 1--6 Mm ({\it
left}) and 8--20 Mm ({\it right}) in the region of the flux
emergence of AR 10488. The units of ${\rm div}V_h$ are $3\cdot 10^{-7}$ s$^{-1}$.}
    \label{Fig13}
\end{figure}

\begin{figure}
\begin{center}
\includegraphics[width=\linewidth]{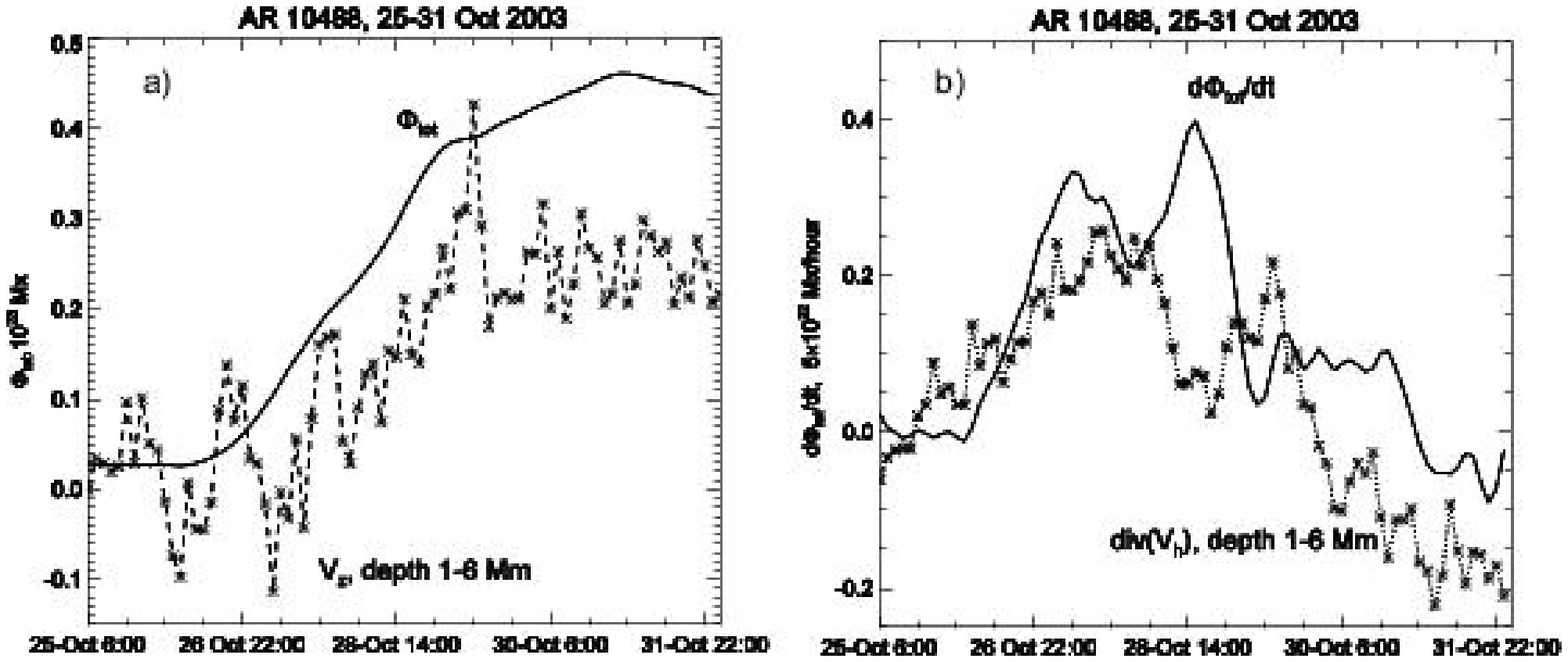}
 \end{center}
   \caption{{\it a}) The evolution of the total unsigned photospheric magnetic
flux (solid curve) and the mean vertical velocity in km/s (dotted
curve with stars) at the depth of 1--6 Mm in the region of the flux
emergence of AR 10488. The negative velocity corresponds to upflows,
and the positive velocity corresponds to downflows. {\it b}) The corresponding
changes of the total emerging flux rate and the mean divergence of the
horizontal flow components.}
    \label{Fig14}
\end{figure}
The helioseismology measurements of subsurface flows are obtained
from the reciprocal travel times, and generally, are less affected
by various kind of uncertainties. They may provide better indicators
of the development of active region structures inside the Sun.
Figure~\ref{Fig12} shows six flow maps at the depth of about 2 Mm
for various stages of evolution of the active region, NOAA 10488, before the
emergence, during the initial emergence, and during the developed
state. The background gray-scale maps show the corresponding
magnetograms.

Prior to the emergence, the maps do not show any specific flow
pattern that would indicate development of a large magnetic
structure below the surface, except, perhaps, a small shearing flow
feature, which appeared near the first magnetic field signal in the
center of Figure~\ref{Fig12}a-b. During the next 8 hours
(Fig.~\ref{Fig12}c), this feature disappears, and a
ring-like magnetic field structure is formed. Within this structure
the flows are clearly suppressed, and they remain suppressed during
further evolution. Also, at the same type a diverging flow pattern
starts developing at the boundaries of the magnetic structures. This
pattern is consistent with the expectation that the emerging
magnetic structure pulls plasma outside. The divergent flow field
becomes stronger as the active region grows (Fig.~\ref{Fig12}
e), but later, it is replaced by a converging flow pattern around
the sunspots (Fig.~\ref{Fig12}f), which was previously observed beneath sunspots
\citep*{Zhao:2001}.

The strength of the divergent flows is obviously related to the
development of active regions, and, perhaps, may be even used for predicting
 their future evolution. The time evolution of the mean horizontal
divergence in the two depth intervals and the photospheric flux is
shown in Figure~\ref{Fig13}. It is quite clear that the divergence
at the depth 1--6 Mm started to grow before the magnetic flux,
reached maximum in the middle of the flux growth phase, and then was
replaced by converging flows. At greater depths, 8--20 Mm
(Fig.~\ref{Fig13} right), the horizontal flow behavior is not very
clear, probably because of higher noise, or because the flow pattern
is not as well organized as in the subsurface (6 Mm deep) layer.
Perhaps the most significant feature at this depth is the formation
of a divergent flow pattern approximately at the time of the
formation of convergent flows in the upper subphotospheric layer.

One would expect that during the emergence the plasma is not only
pushed outside the magnetic field area but also upward,
particularly, in the upper layers. Figure~\ref{Fig14}a shows the
evolution of the mean vertical flow below the active region at the
depth 1--6 Mm. Indeed, upflows dominate at the very beginning of the
magnetic flux emergence. However, the signal fluctuates, probably reflecting
a complicated structure of the vertical flows. After the emergence
phase the vertical flow pattern is dominated by downflows, which are
organized around the sunspots.

It seems that the horizontal divergence of subsurface flows is the
most sensitive characteristic of the emerging magnetic flux. The
divergent flows appear before the initial flux emergence, and
continue to evolve in correlation with the magnetic flux.
Figure~\ref{Fig14}b shows a comparison of the mean horizontal
divergence and the total magnetic rate. Evidently, there were two or
three peaks of the magnetic emergence rate. The flow divergence
shows two peaks, which are shifted relative to the flux rate. It is
unclear whether these peaks precede or follow the magnetic flux
emergence events. Obviously, this relationship requires further
investigation. Similar flow patterns have been studied by \citet{Komm:2008}
using data from the GONG network and the ring-diagram method of
local helioseismology. The initial results are quite encouraging and show
the potential of the helioseismic diagnostics.

\section{Comparison with Theoretical Models}

\begin{figure}[!ht]
\centerline{\includegraphics[width=1\linewidth]{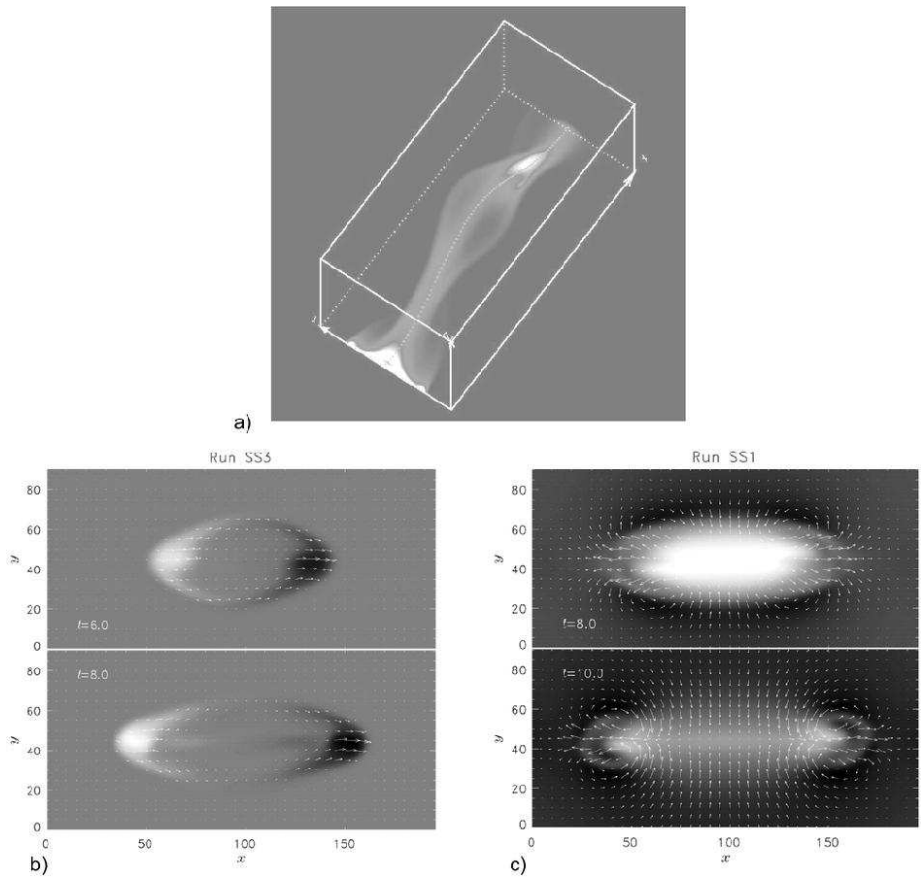}}
\caption{A theoretical MHD model of emerging magnetic flux tube
(Abbett et al. 2000): {\it a}) Volume rendering of the magnetic
field strength. {\it b}) Vector magnetogram images. The gray-scale
background represents the vertical component of the magnetic field
(positive values are indicated by the light regions), and the arrows
represent the transverse components. {\it c}) Velocity field for the
same slice. The gray-scale background represents the vertical
component of the velocity (light regions indicate upflows), and the
arrows represent the transverse components of the velocity
field.}\label{Fig15}
\end{figure}

The time-distance helioseismology measurements provide new
information about the structure and dynamics of emerging active
regions. In this paper, we presented in detail the results for a
large active region, NOAA 10488, which emerged in October 2003. The
results show that the formation of the active region takes about 5
days. During this period the total magnetic flux and the
corresponding subsurface wave-speed perturbation grow mostly
monotonically. However, the magnetic flux rate reveals two or three
peaks of intensive flux emergence; each is about one day long. It
appears that the active region is formed by multiple magnetic flux
emergence events. The initial magnetic flux emerged very rapidly
without any significant perturbation of the Sun's thermodynamic
structure or flow field in the place of emergence. There seems to
be a short lead in the growth of the subsurface wave speed
perturbation relative to the mean magnetic flux, but this
relationship is rather uncertain. A localized shearing flow seems
to be formed few hours before the initial flux emergence, and then
disappears soon after the emergence. The active region has a
elliptical shape, with the magnetic field concentrated at the
boundaries. The plasma flows are suppressed inside this structure.
In the outer region, the plasma flows are dominated by divergent
flow, driven by the expanding magnetic structures. The flow
divergence at the depth of 1--6 Mm shows a few-hour lead relative to
the magnetic flux. It grows during the emergence phase until
approximately the mid-point of the flux growth curve. After this,
the divergence is sharply reduced and then is replaced by
predominantly converging flows around sunspots. Approximately, at
the same time a divergent flow pattern is formed in the deeper
interior (depth 8--20 Mm). The vertical flow pattern is quite
complicated. The results do not show strong upflows prior to the
emergence, as one might expect. In general, the vertical flow
pattern is highly intermittent. There is an evidence of predominant
upflows during the initial stage of emergence, but after this the
mean flow beneath the active region is directed downwards. It seems
that there is an interesting correlation with some time lag between
the flow divergence and the flux emergence rate. However, at this
stage it is unclear whether the changes in the flow divergence
precede or follow the flux rate.

It is interesting that the magnetic structure of this active region,
in particular, its elliptical shape is very similar to a model of
emerging magnetic flux tube of \citet*{Abbett:2000}
(Fig.~\ref{Fig15}). The model also predicts a divergent flow
pattern similar to the observed one. However, the strong upflow at
the beginning of emergence (Fig.~\ref{Fig15}c, top panel)
is not detected in our observations. Thus, the process of emergence
and formation of active regions requires further observational and
theoretical studies.

\section{Discussion and Future Perspectives}

The recent observations and modeling reveal some interesting features
of the properties of emerging magnetic flux and associated dynamics on
the solar surface and in the upper convection zone. In particular,
the new statistical study of the variations of the tilt angle
of bipolar magnetic regions during the flux emergence questions the
current paradigm that the magnetic flux emerging on the solar surface
represents large-scale magnetic flux ropes ($\Omega$-loops)
rising from the bottom of the convection zone. The flux rope models
predict that the tilt angle is a result of the Coriolis effect acting
on a plasma flow inside the flux tube, and thus the tilt should depend
on latitude, the amount of magnetic flux and relax after
the emergence when the Coriolis force vanishes. The observations indeed
show the predicted latitudinal dependence (the Joy's law) and indicate that the
tilt is formed below the surface. However, there is no evidence of
the dependence on the amount of magnetic flux and no evidence for the relaxation
of the tilt angle towards the East-West direction. Contrary, the tilt angle
tends to relax to the Joy's law value. Perhaps, the Joy's law reflects
not the dynamics of the rising flux tubes but the orientation of the
toroidal magnetic field lines below the surface as suggested by \citet{Babcock:1961}.

The observations of the surface flows from SOHO/MDI prior and during the emergence
of a large active regions, AR 10488, in October 2003, show strong localized
vertical flows just prior the flux emergence and during the initial stage.
It is curious that the direction of the flows, namely, an upflow in the area of
the leading polarity and a downflow in the following polarity, is consistent
with the predictions of the rising flux rope theories. However,
observations of some other active regions do not show this \citep{Pevtsov:2006}.
Also, the data do not show large-scale flow patterns on the surface, which
would indicate emergence of a large flux-rope structure.

The local helioseismology results obtained by both, the time-distance and
ring-diagram techniques, show large-scale outflows beneath the surface
during most of the emergence phase, and also formation of converging flows
around the magnetic structure of sunspots. However, the structure of the
vertical flows remains unclear. There is an indication of upflows mixed
with downflows at the beginning of emergence, but then the downflows dominate.
In the case of AR 10488, there were two or three major flux emergence events.
The photospheric magnetic flux rate and subsurface flow divergence show two
or three peaks, which are not in phase, but it is unclear if the flux rate
precedes the variation of the flow divergence or follows it.

From the observations it is obvious that
the multiple flux emergence events over several days plays important role in the formations
and maintaining the magnetic structure of the large active region. This
reminds the idea of a common `nest' in the deep interior \citep{Castenmiller:1986}.
However, such nests have not been found in the helioseismic
images of the subphotospheric magnetosonic wave speed variations, which are currently obtained
 up to the depth of 40-50 Mm. The wave speed images reveal that
the emerging magnetic flux structures travel very fast in the upper convection zone,
with a speed of at least 1 km/s. This makes very difficult the detection of these structures
before the magnetic field becomes visible on the surface. Thus, it is difficult
to use the helioseismology measurements for advanced predictions of emerging
active regions. However, it should be possible to use the measurements of both,
the wave speed variations and flow velocities, for predicting the growth and
and decay of active regions and, perhaps, the complexity of their magnetic
structure. This task will require a substantial statistical analysis of emerging
active regions by methods of local helioseismology.

Thus, despite the significant new information from helioseismology and magnetography
the main questions formulated in Introduction about the origin and physical properties of the emerging magnetic remain unanswered. The recent results from the SOHO spacecraft
and GONG network show that for further investigations it is necessary to improve
the local helioseismology techniques, extending their coverage into the deep
convection zone, carry out statistical studies using uninterrupted solar oscillation
data (such as will be available from the Solar Dynamics Observatory mission), and
also develop realistic MHD numerical simulations for understanding the physics
of magnetic structures in the turbulent convection zone and for supporting the
helioseismology observations.

\begin{acknowledgements} I wish to thank all the organizers and
participants of this workshop, and, in particular, Dr Andre Balogh.
This work was supported by the International Space
Science Institute (Bern).
\end{acknowledgements}

\bibliographystyle{ssrv}

\end{document}